\documentclass{article}
\usepackage[T1]{fontenc}
\usepackage[latin9]{inputenc}
\usepackage{color}
\usepackage{graphicx}
\begin{document}

\title{\textcolor{black}{Galaxy rotations from quantised inertia and visible
matter only.}}

\author{\textcolor{black}{M.E. McCulloch}%
\thanks{\textcolor{black}{Plymouth University, Plymouth, PL4 8AA. mike.mcculloch@plymouth.ac.uk}%
}}
\maketitle
\begin{abstract}
\textcolor{black}{It is shown here that a model for inertial mass,
called quantised inertia, or MiHsC (Modified inertia by a Hubble-scale
Casimir effect) predicts the rotational acceleration of the 153 good
quality galaxies in the SPARC dataset (2016 AJ 152 157), with a large
range of scales and mass, from just their visible baryonic matter,
the speed of light and the co-moving diameter of the observable universe.
No dark matter is needed. The performance of quantised inertia is
comparable to that of MoND, yet it needs no adjustable parameter.
As a further critical test, quantised inertia uniquely predicts a
specific increase in the galaxy rotation anomaly at higher redshifts.
This test is now becoming possible and new data shows that galaxy
rotational accelerations do increase with redshift in the predicted
manner, at least up to Z=2.2.}
\end{abstract}

\section{\textcolor{black}{Introduction}}

\textcolor{black}{It has been well known since van Oort (1932), Zwicky
(1933) and Rubin et al. (1980), that galaxies rotate far too fast
to be gravitationally stable. The usual solution for this is to add
dark matter to the galactic haloes to hold stars in with more gravitational
force. This solution is ad hoc since it has to be added to different
galaxies in different amounts. It is also difficult to falsify so
it is therefore unsatisfying, and it has recently been shown by McGaugh
et al. (2016) that the acceleration of stars in galaxies is correlated
with the distribution of the visible matter only, which implies there
is no dark matter.}

\textcolor{black}{One alternative to dark matter is MoND (Modified
Newtonian Dynamics) (Milgrom, 1983) in which either the gravitational
force on, or the inertial mass of, orbiting stars is changed for very
low accelerations. MoND is an empirical hypothesis that has no physical
model and relies on its adjustable parameter (a0) which is fitted
to the data by hand, which is unsatisfactory since no justification
is given for this parameter.}

\textcolor{black}{Fulling (1973), Davies (1975) and Unruh (1976) proposed
that when an object accelerates it perceives Unruh radiation, a dynamical
equivalent of Hawking radiation (Hawking, 1974), and Unruh radiation
may now have been seen in experiments (Smolyaninov, 2008).}

\textcolor{black}{McCulloch (2007, 2013) has proposed a new model
for inertia that assumes that when an object accelerates, say, to
the right, an information horizon forms to its left and it perceives
Unruh radiation which is also suppressed by the horizon on the left.
Therefore there is a radiation imbalance, and net Unruh radiation
pressure that pushes the object back against its initial acceleration,
predicting standard inertia (McCulloch, 2013, Gine and McCulloch,
2016). Furthermore, this model predicts that some of the Unruh radiation
will also be suppressed, this time isotropically, by the distant cosmic
horizon which will make this mechanism less efficient, reducing inertial
mass in a new way, especially for very low accelerations for which
Unruh waves are very long (McCulloch, 2007). The complete model, called
MiHsC (Modified inertia by a Hubble-scale Casimir effect) or quantised
inertia modifies the standard inertial mass ($m$) as follows:}

\textcolor{black}{
\begin{equation}
m_{i}=m\left(1-\frac{2c^{2}}{|a|\Theta}\right)
\end{equation}
}

\textcolor{black}{where $c$ is the speed of light, $\Theta$ is the
Hubble diameter and $|a|$ is the magnitude of the acceleration of
the object relative to surrounding matter. Eq. 1 predicts that for
terrestrial accelerations (eg: $9.8m/s^{2}$) the second term in the
bracket is tiny and standard inertia is recovered, but in environments
where the mutual acceleration is of order $10^{-10}m/s^{2}$, for
example at the edges of galaxies or in dwarf galaxies, the second
term becomes larger and the inertial mass decreases in a new way.
This modification does not affect equivalence principle tests using
torsion balances or free fall since the predicted inertial change
is independent of the mass.}

\textcolor{black}{In this way quantised inertia explains galaxy rotation
without the need for dark matter (McCulloch, 2012, 2017) because it
reduces the inertial mass of outlying stars and allows them to be
bound even by the gravity from visible matter. It also explains the
recently observed cosmic acceleration (McCulloch, 2010) and the experimental
tests on the emdrive (McCulloch, 2015).}

\textcolor{black}{Recently, McGaugh et al. (2016) analysed 153 galaxies
taken from the SPARCs database across a large range of scales and
showed that the actual acceleration of the stars within them, as determined
from the stars' observed motion, was correlated only with the acceleration
that would be expected given the visible matter in the galaxy. This
result has now also been shown to apply to elliptical galaxies (Lelli
et al., 2017). As mentioned above, these results argue against the
existence of dark matter. McGaugh et al. (2016) also found that the
relationship between the observed acceleration ($a_{obs}$) and that
expected from the visible or baryonic matter ($a_{bar}$) could be
described quite well by the function}

\textcolor{black}{
\begin{equation}
a_{obs}=\frac{a_{bar}}{1-e^{-\sqrt{a_{bar}/a_{0}}}}
\end{equation}
}

\textcolor{black}{It has been implied that this function can be obtained
from some versions of Modified Newtonian Dynamics (MoND) of Milgrom
(1983), though this empirical model needs to be adjusted to fit, and
has no supporting physical model.}

\textcolor{black}{In this paper, it will be shown that quantised inertia
can predict the new galaxy data presented by McGaugh et al. (2016)
without any adjustable parameters, simply from the visible matter,
the speed of light and the co-moving diameter of the observable universe.
It is also shown that quantised inertia can be tested because it uniquely
predicts a significant change in the galactic acceleration relation
with redshift.}

\section{\textcolor{black}{Method}}

\textcolor{black}{To briefly recapitulate McCulloch (2012), we start
with Newton's gravity and second laws, for a star of mass $m$ orbiting
a galaxy of mass $M$ at radius $r$ as follows}

\textcolor{black}{
\begin{equation}
F=m_{i}a=\frac{GMm}{r^{2}}
\end{equation}
}

\textcolor{black}{Replacing $m_{i}$ using quantised inertia, Eq.
1, we get}

\textcolor{black}{
\begin{equation}
\left(1-\frac{2c^{2}}{|a|\Theta}\right)a=\frac{GM}{r^{2}}
\end{equation}
}

\textcolor{black}{Splitting the acceleration $|a|$ up into a slowly
varying (rotational) part $a=v^{2}/r$ and a variable part $a'$ due
to inhomogeneities in the matter distribution, gives}

\textcolor{black}{
\begin{equation}
\left(|a|+|a'|-\frac{2c^{2}}{\Theta}\right)a=\frac{GM(|a|+|a'|)}{r^{2}}
\end{equation}
}

\textcolor{black}{At the edge of a galaxy, $|a|$ becomes small, so
the acceleration must be maintained above the minimum acceleration
allowed in quantised inertia (McCulloch, 2007) by the value of $a'$,
and so $a'=2c^{2}/\Theta$. Therefore the second and third terms cancel.
In this way, McCulloch (2012) derived the following formula}

\textcolor{black}{
\begin{equation}
a^{2}=\frac{GM(|a|+|a'|)}{r^{2}}
\end{equation}
}

\textcolor{black}{The }\textit{\textcolor{black}{a}}\textcolor{black}{{}
on the left hand side can be called the predicted total acceleration
$a_{pred}$. The factor $GM/r^{2}$ and the $|a|$ on the right hand
side can be replaced with $a_{bar}$: the baryonic, standard model,
acceleration. Assuming, again, that at a galaxy's edge the residual
acceleration $a'=2c^{2}/\Theta$ since accelerations cannot fall below
this minimum in quantised inertia, then we get}

\textcolor{black}{
\begin{equation}
a_{pred}^{2}=a_{bar}^{2}+a_{bar}\frac{2c^{2}}{\Theta}
\end{equation}
}

\textcolor{black}{which leads to the formula}

\textcolor{black}{
\begin{equation}
a_{pred}=a_{bar}\sqrt{1+\frac{2c^{2}}{a_{bar}\Theta}}
\end{equation}
}

\textcolor{black}{In the next section this prediction is compared
with the raw SPARC data collated by McGaugh et al. (2016).}

\section{\textcolor{black}{Results}}

\textcolor{black}{Figure 1 shows the log of the expected (Newtonian)
stellar acceleration, predicted from the visible mass, on the horizonal
axis and the log of the observed accelerations, observed from stellar
motions, on the vertical axis. The expected result from standard physics
is the diagonal dotted line. The grey squares show the observed accelerations
from the binned data obtained from McGaugh et al. (2016) (pers. comm.).
The size of the squares show the rms error in each bin. For very low
accelerations (on the left) the data lifts above the dotted line,so
that the observed accelerations are much higher than those expected
from the standard Newtonian (or general relativistic) model. This
is the well-known galaxy rotation problem.}

\textcolor{black}{The black line shows the prediction of one of the
variations of MoND (Modified Newonian Dynamics) of Milgrom (1983).
MoND agrees with the data, but it has been fitted to galaxy data using
its adjustable parameter (the value used here was $a_{0}=1.2\times10^{-10}m/s^{2}$)
so the agreement is not so remarkable.}

\textcolor{black}{The prediction of quantised inertia for a redshift
of zero, the present epoch, is shown by the dashed line, and it also
fits the data for the present epoch within the error bars, and this
agreement requires no adjustment at all. The curve is predicted, by
Eq. 8 above, and therefore uses only the visible matter, the speed
of light (c) and the co-moving cosmic diameter ($\Theta$), which
is assumed to be 93 billion light years or $\Theta=8.8\times10^{26}m$
and is the cosmic diameter at the present time following Bars and
Terning (2009) (see page 27). Note that this is different from the
value of $\Theta=2.6\times10^{26}m$ used in McCulloch (2007, 2012)
which represents the cosmic diameter at the epoch when the light from
the galaxies was emitted.}

\section{\textcolor{black}{Discussion}}

\textcolor{black}{It is always useful to suggest a test, by predicting
something unique that has not yet been observed. Unlike MoND, which
uses a constant and pre-set parameter $a_{0}$, quantised inertia
relies on the value of $2c^{2}/\Theta$. This depends on the co-moving
size of the cosmos $\Theta$ which increases with time (or decreases
with time into the past) and so quantised inertia predicts a change
in galaxy rotation with time. This can be seen in Eq. 8 which depends
on the cosmic diameter $\Theta$ which was smaller in the distant
past and, assuming a linear expansion of the cosmos with time, depends
on the redshift (Z) as follows}

\textcolor{black}{
\begin{equation}
\Theta_{atZ}=\frac{\Theta_{now}}{1+z}
\end{equation}
}

\textcolor{black}{Therefore, quantised inertia predicts that the acceleration
relation in Fig. 1 should show a dependence on redshift. At higher
$z$ (further back in cosmic time) the galaxy rotation problem should
be more obvious (everything else, such as galaxy evolution, being
taken care of) so that equal-mass galaxies observed in the distant
past should spin faster. This is illustrated in Fig. 1 with the dashed
line which represents the predition of quantised inertia for $z=0$,
the present epoch, the longer-dashed line which shows the prediction
for $z=1$ (when the cosmos was half its present size) and the dot-dashed
curve which shows the prediction for $z=2$. This prediction is unique
to quantised inertia, and the data is now becoming available to test
this prediction. For example, Thomas et al. (2013) looked at galaxies
at different redshifts from $z=0$ to $z=2$ from the Sloan Digital
Sky Survey / Baryonic Oscillation Spectroscopic Survey SDSS/BOSS collaboration,
and showed that higher redshift galaxies have higher velocity dispersions
(see their Fig. 6) but better data is needed to confirm this.}

\textcolor{black}{A more common way of looking at this is to consider
the mass required to produce a particular rotation speed. The required
mass can be derived from quantised inertia as follows. At a galaxy's
edge, since the rotational acceleration (|a|) is so slow, Eq. 6 can
be rewritten as}

\textcolor{black}{
\begin{equation}
a^{2}=\frac{2GMc^{2}}{\Theta r^{2}}
\end{equation}
}

\textcolor{black}{and replacing $a^{2}$ using $v^{4}/r^{2}$ we get}

\textcolor{black}{
\begin{equation}
v^{4}=\frac{2GMc^{2}}{\Theta}
\end{equation}
}

\textcolor{black}{This is the Tully-Fisher relation predicted by quantised
inertia, which now varies with time since $\Theta$ was smaller in
the past. Using Eq. 9 to take account of this evolution, we get}

\textcolor{black}{
\begin{equation}
v^{4}=\frac{2GMc^{2}(1+Z)}{\Theta_{now}}
\end{equation}
}

\textcolor{black}{So that the amount of mass associated with a rotation
speed of v is given by}

\textcolor{black}{
\begin{equation}
M=\frac{v^{4}\Theta_{now}}{2Gc^{2}(1+Z)}
\end{equation}
}

\textcolor{black}{Recently, Ubler et al. (2017) (submitted to ApJ)
found that at redshifts of Z=0.9 the amount of mass associated with
a specific rotation speed is reduced by between -0.38 and -0.47 dex
($dex(x)=10^{x}$, see their Figure 7) and at Z=2.3 the reduction
in mass is between -0.2 and -0.47 dex. Quantised inertia (Eq. 13)
predicts a reduction in mass of -0.28 and -0.52 dex respectively.}

\textcolor{black}{Probably the best source of data on this to date
is Genzel et al. (2017) who looked at six massive galaxies at high
redshifts, between Z=0.854 and Z=2.383, and also showed that their
rotation speed increased at higher redshifts. Another way to model
this with quantised inertia is to note that it precludes accelerations
below $2c^{2}/\Theta$. Figure 2 shows along the x axis the observed
acceleration of the galaxies (calculated from their half-light radii
and their velocity dispersion) and along the y axis the minimum acceleration
allowed by quantised inertia. The six black squares show the comparisons
for the six galaxies, and the adjacent numbers indicate their redshifts.
In both the observations and the predictions from quantised inertia,
the galactic accelerations increase with redshift. The predictions
show the same tendency as the observations (the squares are close
to the line of agreement), but they are between 9\% and 19\% higher
for the four lower redshift galaxies. This difference can be accounted
for by uncertainties in the cosmic expansion model used to determine
$\Theta$. Agreement is worse for the two highest redshift galaxies,
which may be expected to have a larger uncertainty.}

\section{\textcolor{black}{Conclusions}}

\textcolor{black}{A new model for inertia (called quantised inertia
or MiHsC) predicts the observed rotational accelerations of the 153
galaxies in the recent SPARC dataset simply from their visible matter,
the speed of light and the co-moving diameter of the cosmos (Figure
1), without dark matter or any adjustable parameters.}

\textcolor{black}{As a test, quantised inertia uniquely predicts a
significant increase in the galaxy rotation anomaly at higher redshifts
and this is supported by recent data, at least up to a redshift of
Z=2.2.}

\section*{\textcolor{black}{Acknowledgements}}

\textcolor{black}{Many thanks to S.S. McGaugh for making available
the binned SPARC data, and R. Ludwick, T. Short, Magnus Ihse Bursie
and J.A.M. Lizcano and an anonymous reviewer for advice.}

\section*{\textcolor{black}{References}}

\textcolor{black}{Bars, Itzhak and J. Terning, 2009. Extra dimensions
in space and time. Springer, pp. 27-.}

\textcolor{black}{Davies, P.C.W., 1975. J. Phys. A., 8, 609.}

\textcolor{black}{Fulling, S.A., 1973. Phys. Rev. D., 7, 2850.}

\textcolor{black}{Genzel R.et al., 2017. Strongly barylon-dominated
disk galaxies at the peak of galaxy formation ten billion years ago.
Nature, 543, 397-401.}

\textcolor{black}{Hawking, S., 1974. Nature, 248, 30.}

\textcolor{black}{Lelli, F., S.S. McGaugh, J.M. Schombert, M.S. Pawlowsky,
2017. One law to trule them all: the radial acceleration relation
of galaxies. ApJ, 836, 152.}

\textcolor{black}{McCulloch, M.E., 2007. Modelling the Pioneer anomaly
as modified inertia. MNRAS, 376, 338-342.}

McCulloch, M.E., 2010. Minimum accelerations from quantised inertia.
EPL, 90, 29001.

\textcolor{black}{McCulloch, M.E., 2012. Testing quantised inertia
on galactic scales. ApSS, 342, 575-578.}

\textcolor{black}{McCulloch, M.E., 2013. Inertia from an asymmetric
Casimir effect. EPL, 101, 59001.}

\textcolor{black}{McCulloch, M.E., 2015. Testing quantised inertia
on the emdrive. EPL, 111, 60005.}

\textcolor{black}{McCulloch, M.E., 2017. Low-acceleration dwarf galaxies
as tests of quantised inertia. Astrophys. Space Sci., 362, 57.}

\textcolor{black}{McGaugh, S.S, F. Lelli, J. Schombert, 2016. The
radial acceleration relation in rotationally supported galaxies. Phys.
Rev. Lett.}

\textcolor{black}{Gine and McCulloch, 2016. Inertial mass from Unruh
temperatures. Modern Phys. Lett. A., 31, 1650107.}

\textcolor{black}{McGaugh, S., F. Lelli, J. Schombert, 2016. The radial
acceleration relation in rotationally supported galaxies. Phys. Rev.
Lett., 117, 201101.}

\textcolor{black}{Milgrom, M., 1983. A modification of the Newtonian
dynamics as a possible alternative to the hidden mass hypothesis.
Astrophys. J., 270, 365.}

\textcolor{black}{Oort, J.H., 1932. The force exerted by the stellar
system in the direction perpendicular to the galactic plane and some
related problems. Bull. Astronom. Institutes of the Neth., 6, 249-287.}

\textcolor{black}{Rubin, V., N. Thonnard and W.K. Ford Jr., 1980.
Rotational properties of 21 SC galaxies with a large angle of luminosities
and radiifrom NGC 4605 (R=4kpc) to UGC 2885 (R=122kpc). Astrophys.
J., 238, 471.}

\textcolor{black}{Smolyaninov, I.I., 2008. Photoluminescence from
a gold nanotip in an accelerated reference frame. Phys. Lett. A.,
372, 7043-7045.}

\textcolor{black}{Thomas, D., et al., 2013. Stellar velocity dispersions
and emission line properties o SDSS-III/BOSS galaxies. MNRAS, 431,
2, 1383-1397.}

\textcolor{black}{Uebler, H., et al., 2017. The evolution of the Tully-Fisher
relation between z\textasciitilde{}2.3 and z\textasciitilde{}0.9 with
KMOS. Astrophys. J., Vol. 842, No. 2, 121.}

\textcolor{black}{Unruh, W.G., 1976. Phys. Rev. D., 14, 870.}

\textcolor{black}{Zwicky, F., 1933. Der Rotverschiebung von extragalaktischen
Nebeln. Phys. Acta, 6, 110.}

\section*{\textcolor{black}{Figures}}

\textcolor{black}{\includegraphics[scale=0.65]{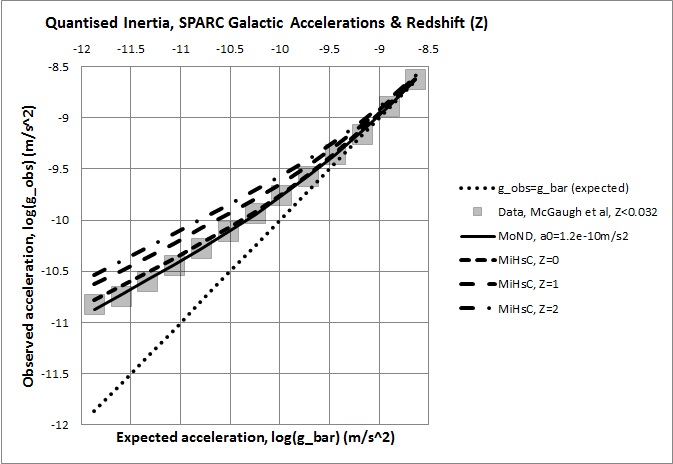}}

\textcolor{black}{Figure 1. The x-axis shows the expected (Newtonian)
acceleration of stars in a galaxy given the visible mass distribution.
The y-axis shows the acceleration observed from the movements of its
stars. The expected Newtonian result is shown by the dotted line.
The data (binned using 3300 data points in 153 seperate galaxies by
McGaugh et al., 2016) is shown by the grey squares and the error bars
are shown by the size of the squares. Both MoND (black line) and quantised
inertia (dashed for redshift Z=0) agree with the observations, but
quantised inertia predicts the data just from the speed of light and
the co-moving diameter of the cosmos (see Eq. 8) whereas MoND and
other solutions like dark matter need arbitrary 'fitting'. Also shown
are the predictions of quantised inertia for earlier galaxies with
redshifts of Z=1 (long-dashed) and Z=2 (dot-dashed).}

\textcolor{black}{\includegraphics[scale=0.8]{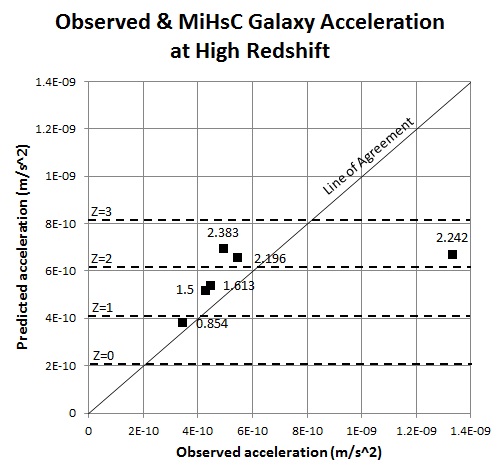}}

\textcolor{black}{Figure 2. The observed acceleration of the six galaxies
(on the x axis) from Genzel et al. (2017) plotted against the minimum
acceleration allowed by quantised inertia at that epoch on the y axis,
which is $2c^{2}/\Theta$ where $\Theta=\Theta_{now}/1+Z$. The redshift,
Z, is shown as a label against each data point and as the horizontal
dahed lines. Both the observed and predicted accelerations rise with
redshift. The predicted accelerations are close to those observed
(close to the diagonal line) except for the two highest redshift galaxies.}
\end{document}